\documentstyle[amsfonts,aps,prl,epsf]{revtex}

\newcommand{\beq}{\begin{equation}}
\newcommand{\eeq}{\end{equation}}
\newcommand{\beqa}{\begin{eqnarray}}
\newcommand{\eeqa}{\end{eqnarray}}
\newcommand{\non}{\nonumber}
\newcommand{\opsi}{\overline{\psi}}

\begin{document}

\draft

\title{Index Distribution of Random Matrices with an Application to
Disordered Systems}

\author{
Andrea Cavagna\thanks{E-mail: a.cavagna1@physics.ox.ac.uk.},
Juan P. Garrahan\thanks{E-mail: j.garrahan1@physics.ox.ac.uk.}
and Irene Giardina\thanks{E-mail: i.giardina1@physics.ox.ac.uk.} }

\address{
Theoretical Physics, University of Oxford \\
1 Keble Road, Oxford, OX1 3NP, UK}

\date{\today}

\maketitle

\begin{abstract}

We compute the distribution of the number of negative
eigenvalues (the index) for an ensemble of Gaussian 
random matrices, by means of the replica method. 
This calculation has important applications in the 
context of statistical mechanics of disordered systems, 
where the second derivative of the potential energy (the 
Hessian) is a random matrix whose negative 
eigenvalues measure the degree of instability of the 
energy surface.
An analysis of the probability distribution of the 
Hessian index is therefore relevant for a geometric 
characterization of the energy landscape in disordered systems.
The approach we use here is particularly suitable for 
this purpose, since it addresses the problem without 
any {\it a priori} assumption on the random matrix ensemble 
and can be naturally extended to more realistic,
non-Gaussian distributions.

\end{abstract}

\pacs{PACS numbers: 75.10.Nr, 61.43.Fs, 64.70.Pf, 61.20.Lc}

\section{Introduction}

The importance of random matrix theory (RMT) can hardly be overstated.
Since its initial development by Wigner and Dyson 
to deal with the spectrum of many-body quantum systems \cite{wigner,dyson}, 
it has found 
applications in areas of  physics as diverse as disordered systems, 
chaos, and quantum gravity, to name just a few \cite{mehta,review,fgz,physrep}. 
Most of the time RMT has been used as a very powerful tool for the 
study of the energy-level fluctuations of quantum systems. 
In this case the matrix that RMT is modeling is of course 
the quantum Hamiltonian of the system.

However, there is a different context where RMT can be very useful, 
namely the study of the statistical properties of classical disordered 
systems. By disordered systems we mean not only those cases 
where quenched disorder is directly present in the Hamiltonian, as in 
spin-glasses, random field models or neural networks, 
but also systems
whose physical behaviour at low temperatures is heavily influenced by
the self-induced disorder of their typical configurations, as, for example,
supercooled liquids and structural glasses. 
In all these systems the properties of the energy landscape, or energy
surface, are known to be far from trivial. In particular, the presence of 
many local minima of the potential energy is one of the most distinctive
features of this class of systems \cite{spin-glass,supercooled}. 
An obvious consequence of this fact 
is that the energy surface displays many extensive regions with unstable 
negative curvature and therefore has very non-trivial stability 
properties \cite{laloux,noiselle}. 
In this context a key object becomes the matrix of the second derivatives
of the Hamiltonian, normally called Hessian, which encodes all the 
stability attributes of the energy landscape. 

The study of the statistical properties of the Hessian has been 
an important issue both in the theory of mean-field spin-glasses 
and in liquid theory. 
In the former case it is often possible to analyze the Hessian in the 
stationary points of the free-energy, having therefore important information 
on the shape and stability of the thermodynamic states \cite{spin-glass}. 
In liquids, on the other hand, the Hessian of the potential energy is the 
key object in the context of the instantaneous normal modes approach 
\cite{stratt,keyes}, where the average spectrum of the Hessian 
is directly connected to many physical 
observables of the system. In particular, it has been argued 
that there exists a deep relation between the diffusion properties
of a liquid and the negative unstable eigenvalues of the 
average Hessian \cite{keyes}.

It is evident that in the above context
an application of RMT to the study of the statistical
properties of the Hessian  can be potentially very
useful. An important remark is the following: the Hessian is a matrix 
which in general depends on the configuration of the system and possibly also
on the quenched disorder, when this is present. The basic
idea is to derive from the distribution of the configurations 
and from the distribution of the disorder an effective 
probability distribution for the Hessian, which can then be studied 
in the context of RMT (a recent example of this strategy can be found 
in \cite{noi-inm}). 

Besides, it is clear from the former discussion that an important issue 
is the analysis of the negative eigenvalues of the Hessian, since their
presence is related to regions of unstable negative curvature of the
energy surface and thus possibly to the boundaries of different basins
of attractions in the phase space. In particular, the number of negative
eigenvalues of the Hessian, called the {\it index}, is the first and easiest
measure of instability. As a consequence, all the tools devised for the
investigation of the index in RMT are particularly relevant
in the context of statistical mechanics of disordered systems.
The average value of the index is trivially related to the average 
spectrum of the Hessian by a simple integration. On the other hand, a 
more interesting and less trivial quantity is the {\it probability 
distribution} of the index. 
Indeed, while the average index gives a measure of the overall
degree of instability of the energy surface, the knowledge of the 
fluctuations of the index around its average value allows a more 
profound  and complete geometric description the energy landscape.

In this paper we compute the probability distribution of the index 
for an ensemble of Gaussian random matrices with a diagonal shift. 
This ensemble provides the simplest possible model for the Hessian 
of a disordered system at a given energy and represents the ideal 
context where to develop the technical aspects of this kind of 
computation. Moreover, in the Gaussian context we are able to 
give non-trivial physical interpretations of our results.

In order to compute the index distribution we use a fermionic replica method. 
In the past the replica method has been applied 
to recover standard results in RMT, with variable success.
Recently the interest of the community has focused again on this
method \cite{meka1,meka2,yl} and some indications of the mathematical 
consistency of the method have been provided, even if some strong criticisms 
still persist \cite{zirn}. The present computation offers an interesting 
example where the replica method can be applied to obtain exact results 
which are not easily available in the standard RMT literature.

There is also another important reason for using the replica method in
the computation of the index, which is related to the physical 
relevance of the Hessian discussed above. 
As we have seen, RMT can be used once an effective
probability distribution for the Hessian has been worked out from the 
distribution of the configurations and from the distribution of the 
quenched disorder. This
effective distribution will not be Gaussian in general (unless we consider 
some very particular models) and typically it will not belong to the standard 
ensembles considered by ordinary RMT. 
By means of the replica method we have in principle no need to assume any
specific form of the distribution.

The paper is organized as follows. 
In Sec. II we compute the average determinant for matrices
of the Gaussian Orthogonal Ensemble as a warm-up exercise to fix
notation and ideas. We then proceed in Sec. III to the main part of the paper,
where we calculate the average index distribution by means of the replica 
method,
in the limit of large matrices.  In Sec. IV we apply the previous analysis
to the specific case of a mean-field spin-glass model, where the Hessian 
is exactly a Gaussian random matrix. Finally in Sec. V we discuss the  
general relevance of our results and state our conclusions . 
Technical details of the calculation and the contribution of replica
symmetry broken solutions are contained in two appendices.

\section{A Preliminary Calculation}

Consider the matrix
\beq
M_{ij} = J_{ij} - E \, \delta_{ij},
	\label{eq:mat}
\eeq
where $J_{ij}$ is an $N$-dimensional real and symmetric 
random matrix with the Gaussian distribution function
\beq
{\cal P}[J] = 2^{-N/2} 
	\left( \frac{N}{\pi} \right)^{N^2/2}	
	\exp\left(-\frac{N}{4} {\rm Tr} \, J^2 \right) .
\label{dist}
\eeq
We have introduced a diagonal shift $E$ in order to mimic what
in general happens in disordered systems, where $M$ represents the 
Hessian of the Hamiltonian. In this context we
expect to find very few negative eigenvalues of $M$ at low 
energies, because of the dominance
of minima at very low energies. This is the effect of the 
shift $E$ in (\ref{eq:mat}) 
and we therefore shall refer in the following to $E$ as
to the {\it energy}. 

The average density of eigenvalues, or spectrum, of $M$ is
defined by
\beq
\rho(\lambda;E) = 
	-  \frac{1}{\pi N}\, {\rm Im} \,
	\overline{ {\rm Tr} \left( \lambda - 
	M + i\, \epsilon \right)^{-1}} 
	=  - \frac{1}{\pi N}\, {\rm Im}\; 
	\frac{\partial}{\partial\lambda}\; \overline{
	\log \det ( \lambda - M + i\, \epsilon) } \  ,
	\label{rho1}
\eeq
where the bar indicates the average over distribution (\ref{dist}).
It is well known that for the Gaussian ensembles 
the spectrum $\rho$ in the limit ${N \to \infty}$ is given by a semi-circle
centered around $\lambda=- E$, that is
\beq
\rho(\lambda;E)= 
	\frac{1}{2\pi} \sqrt{4 - (\lambda+E)^2}  \ , 
	\label{semi}
\eeq
while $\rho$ is zero outside the semi-circle support\cite{mehta}.

In order to fix our notation and to acquire some familiarity
with the method we will use, we compute in 
this section the average determinant of $M$. In general this
is {\it not} a self-averaging quantity, in the sense that fluctuations
around the mean value do not decrease in the limit $N\to\infty$.
The correct object to average is in principle the logarithm of the 
determinant, as it appears in the definition of $\rho$, since this is 
an extensive quantity. 
However, it is a particular property of the Gaussian case 
that the determinant {\it is} self-averaging at the leading order, 
so that the calculation of $\overline{\det M}$ is an interesting and
simple  warm-up exercise for what we want to show later.

We can write the determinant by means of a Gaussian integral 
over $N$-dimensional fermionic vectors $(\opsi,\psi)$
\beq
\det M =
	\int d\opsi \, d\psi \,
	\exp\left[ - \sum_{i,j=1}^N \opsi_i \psi_j 
	\left( J_{ij} - E \delta_{ij} \right)
	\right] \ ,
\eeq
We now average over the symmetric matrix $J_{kl}$
\beq
\overline{\det M} =
	\int d\opsi \, d\psi \, 
	\exp \left( 
	E \, \sum_{i=1}^N \opsi_i \psi_i - 
	\frac{1}{2N} \sum_{i,j=1}^N \opsi_i \psi_i \opsi_j \psi_j
	\right) .
\eeq
To decouple the quartic term in the fermions we perform a
Hubbard-Stratonovich transformation 
\beq
\overline{\det M} = 
	\int d\opsi \, d\psi \, dq \,
	\exp \left(
	E \, \sum_{i=1}^N \opsi_i \psi_i -
	\frac{N}{2} q^2 + i \, q \, \sum_i \opsi_i \psi_i
	\right) , 
\eeq
and after integrating out the fermions we obtain
\beq
\overline{\det M} =
	\int dq \, e^{N S(q,E)} \ ,
	\label{int1}
\eeq
with
\beq
S(q,E) =  -\frac{1}{2} q^2 + \log(-E -iq) \ . 
	\label{azione}
\eeq
This integral can be solved exactly in the limit $N\to\infty$ 
by means of the steepest descent method. The procedure is quite standard 
\cite{bender}, but we briefly summarize it for the sake of clarity. 
In order to calculate integral (\ref{int1}) in the large $N$ limit
we must select a path of integration $\gamma$ in the complex plane,
which satisfies the following conditions:

\vskip 0.3 truecm

\noindent
{\it i)} The integral along $\gamma$ must be equal to the integral 
along the original integration path (in our case the real axis). 

\noindent
{\it ii)} The imaginary part of the action $S(z,E)$ (or phase) 
must be constant along $\gamma$.

\noindent
{\it iii)} The path $\gamma$ must pass through {\it at least} one of 
the saddle points of the action $S(z,E)$.

\vskip 0.3 truecm

The integral along $\gamma$ can then be computed using the Laplace method
\cite{bender} and it is given, at the leading order, by the 
integrand evaluated in the maximum of the real part of $S$ along  
$\gamma$, that is, in the saddle point of the whole action. 
In the case where many maxima lie on $\gamma$, only those with the 
largest real part of $S$ contribute to the total integral.

In our case the action $S$ has two saddle points in the complex plane, 
given by
\beq
q_\pm = \frac{i}{2} E \pm 
	\frac{1}{2} \sqrt{4 - E^2} \ .
	\label{qpm}
\eeq
The regions of constant phase passing through $q_+$ and
$q_-$ are defined by 
\beqa
\gamma_+ : \,\, {\rm Im }\, S(z)&=&{\rm Im }\, S(q_+)\nonumber  \\
\gamma_- : \,\, {\rm Im }\, S(z)&=&{\rm Im }\, S(q_-) \ .
\eeqa
These regions satisfy by definition conditions {\it (ii)} and {\it (iii)}
and thus the correct path of integration $\gamma$ must be built by using 
the different branches of $\gamma_+$ and $\gamma_-$ in such a way to 
satisfy condition {\it (i)}. We can distinguish three different regimes:

\vskip 0.3 truecm 

$\bullet \ E < -2$: 
For these values of the energy the imaginary part of the action is 
the same for the two saddle points. The constant phase region is shown 
in Fig.1: it is clear that there is only one path $\gamma$ satisfying
condition {\it (i)} which can be built by means of the different branches 
of the constant phase region. This path is almost parallel to the real axis
and passes through $q_+$, but {\it not} through $q_-$. Indeed, 
the path parallel to the imaginary axis, which passes through both 
the stationary points, does not conserve the original integral.
The only stationary point contributing to the integral is 
therefore $q_+$  and we have
\beq
\overline{\det{M}} = e^{N S(q_+,E)} = 
	2^{-N} \, \left( |E| - \sqrt{E^2 - 4} \right)^N \,
	e^{ N \, \left( |E| + \sqrt{E^2 - 4} \right)^2 / \, 8} \ .
\eeq
In this energy regime the spectrum $\rho$ has support completely 
contained in the positive semi-axis and we thus expect the average 
determinant to be positive, as it is.
\begin{figure}
\begin{center}
\leavevmode
\epsfxsize=5in
\epsffile{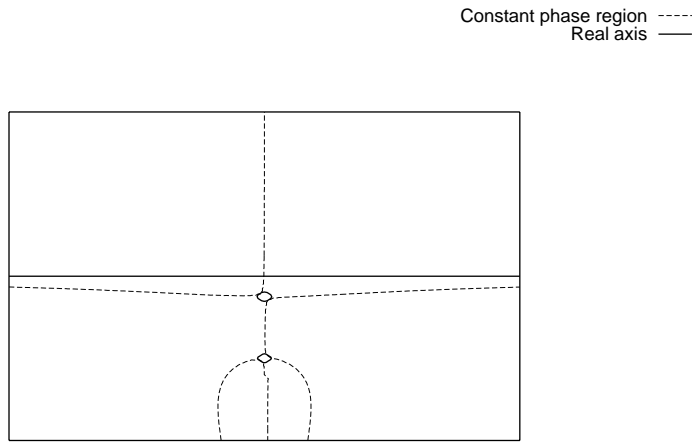}
\caption{$E < -2$: The region of constant phase 
(dashed line) and the real axis (full line). The two small circles indicate
the positions of the two saddle points, $q_+$ (up) and $q_-$ (down). The only 
suitable path of integration $\gamma$ passes just through $q_+$, 
since the original
integral is not conserved on the orthogonal path.  
The case $E>+2$ is specular 
to this one.}
\label{fig1}
\end{center}
\end{figure}

$\bullet \  E > 2$:
The support of the spectrum is
now entirely contained in the negative semi-axis, so we expect 
all eigenvalues of the matrix to be negative. 
In this case the path $\gamma$ passes only through the saddle point
$q_-$, and we thus find for the determinant
\beq
\overline{\det{M}} = e^{N S(q_-,E)} = 
	(-1)^N \, 
	2^{-N} \, \left( |E| - \sqrt{E^2 - 4} \right)^N \,
	e^{ N \, \left( |E| + \sqrt{E^2 - 4} \right)^2 / \, 8} \ ,
\eeq
with the correct prefactor $(-1)^N$ indicating that all eigenvalues are
negative.

$\bullet \ -2 < E < +2$:
In this regime the situation is very different. In Fig.2 we plot
the region of constant phase: the only path $\gamma$ 
which satisfies condition {\it (i)},
passes now through {\it both} the saddle points $q_+$ and $q_-$.
It must be noted that in this case the imaginary part of $S$
is different in $q_+$ and $q_-$, so that actually the global
region of constant phase plotted in Fig.2 is the union of two 
different regions, $\gamma_+$ and $\gamma_-$. On the other hand,
the real part of $S$ is the same in the two stationary points,
and therefore they both contribute to the integral.
We have
\beq
\overline{\det{M}} = 
	e^{N S(q_+,E)} + e^{N S(q_-,E)} =
	(-1)^{N \alpha(E)} \, 
	e^{N (E^2  -  2)/4  + \log 2}  \ , 
\eeq
where
\beq
\alpha(E)=
	\frac{1}{\pi} {\rm arctg}\left(\frac{- \sqrt{4-E^2}}{E}\right)
       +\frac{1}{4\pi} E\sqrt{4-E^2} \ ,
	\label{medio}
\eeq
\begin{figure}
\begin{center}
\leavevmode
\epsfxsize=5in
\epsffile{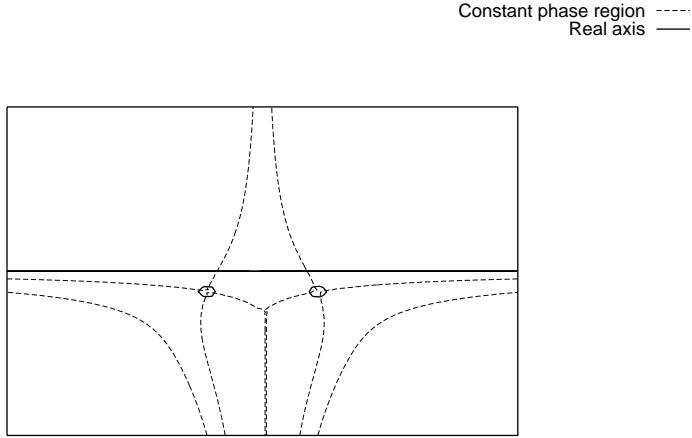}
\caption{$-2<E<2$: The region of constant phase 
(dashed line) and the real axis (full line). The two small circles indicate
the positions of the two saddle points, $q_+$ (right) and $q_-$ (left).
In this case the correct path of integration $\gamma$ passes through 
both the saddle points. }
\label{fig2}
\end{center}
\end{figure}
At these values of the energy the spectrum of $M$ is partly
contained in the negative semi-axis, so that a non-trivial fraction
of the eigenvalues is negative. The interesting point is that the interplay 
between the two saddle points gives rise to the correct sign of the 
determinant. Indeed, it is easy to check that 
$\alpha(E)$ is exactly the mean fraction of negative 
eigenvalues of $M$, that is 
\beq
\alpha(E)=\int_{-\infty}^0 d\lambda \; \rho(\lambda;E) \ .
\eeq

Note that the mechanism we have described above,
given by the interplay between the two saddle points and the paths of 
integration, is crucial in order to obtain the correct result for 
the determinant of $M$.

\section{The index distribution}

The index ${\cal I}_M$ of a matrix $M$, defined as the number of its negative 
eigenvalues, can be computed from the following formula \cite{Jorge}
\beq
{\cal I}_M= \frac{1}{2 \pi i} \lim_{\epsilon\to 0} \ \left[
	\log \det (M - i \epsilon) - 
	\log \det (M + i \epsilon)
	\right] \ .
	\label{index}
\eeq
The meaning of this relation is quite clear: the function 
$f(z) = \log \det (M-z)$ has a cut on the real axis at each eigenvalue 
of $M$, such that by means of the limit in (\ref{index}) we are 
crossing as many cuts as negative eigenvalues are present. 
Besides, this formula can be simply obtained by integrating the 
non-averaged spectrum (\ref{rho1}) from minus infinity up to zero.
In the case we are considering, the index is a function of the
energy $E$ and its average value is given by $N \alpha(E)$ 
(Eq.(\ref{medio})).

We are interested in calculating the average probability distribution
of the index, at a given energy $E$, that is the probability $P(K; E)$
to have a matrix $M$ with index ${\cal I}_M$ equal to $K$, at energy $E$
\beq
P(K; E) = \overline{ \delta( K - {\cal I}_M(E) ) } \ .
	\label{pepe}
\eeq
In the following it will be important to distinguish between the 
{\it extensive} index $K$, which is a positive integer between $0$
and $N$, and the {\it intensive} one $k=K/N$, which takes values in
the continuous interval $[0,1]$, and 
whose probability distribution is
\beq
p(k;E) = \overline{ \delta( k - {\cal I}_M(E)/N ) }= N P(Nk;E) \ .
\eeq
Note that the limit $N\to\infty$ is well defined only for $p(k;E)$.

From Eqs.(\ref{index}) and (\ref{pepe}) we get
\beq
P(K;E) = \frac{1}{2 \pi} \int_{-\infty}^{\infty}
	d\mu \; e^{-i \mu K} \ \overline{G(\mu, E)} \ ,
	\label{pi}
\eeq
where
\beq
G(\mu,E) = {\det}^{\mu/2\pi} (M - i \epsilon) \;
	   {\det}^{-\mu/2\pi}(M + i \epsilon) \ .
	\label{dets}
\eeq
We now make use of the replica method to represent 
the powers of the determinants in $G(\mu,E)$ as analytic continuations
of integer powers
\beq
{\det}^{\pm\mu/2\pi} (M \mp i \epsilon) = 
	\lim_{n_\pm\to\pm\mu/2\pi} 
	{\det}^{n_\pm} (M \mp i \epsilon) \ .
\eeq
By introducing two different sets of $N$-dimensional fermionic 
vectors
$(\bar\chi_\pm^r, \chi_\pm^r)$ with $r=1,\dots,n_\pm$, we can
rewrite the determinants as 
\beq
{\det}^{\pm\mu/2\pi} (M \mp i \epsilon) =
	\lim_{n_\pm\to\pm\mu/2\pi} 
	\int D\bar\chi_\pm^r D\chi_\pm^r 
	\exp \left( - \sum_{r=1}^{n_\pm}
	\bar{\chi}_\pm^r (M \mp i\epsilon) \chi_\pm^r \right) \ ,
	\label{fer}
\eeq
where the sums over the matrix indices $i,j$ are hereafter
always understood. 
We can write everything in a more compact fashion by introducing 
the Grassmann vectors
$(\bar\psi_a,\psi_a)$, with $a = 1, \ldots, (n_+ + n_-)$, defined as
(see also \cite{nomo}) 
\beq 
(\psi_1, \ldots, \psi_{(n_+ + n_-)}) 
	 \equiv 
	(\chi_+^1, \ldots, \chi_+^{n_+}, \chi_-^1, \ldots, \chi_-^{n_-}) \ , 
\eeq
together with the matrix
\beq
\epsilon_{ab} \equiv \mbox{} {\rm diag} (
	\underbrace{\epsilon,\ldots,\epsilon}_{n_+}, 
	\underbrace{-\epsilon,\ldots,-\epsilon}_{n_-}) \ .
\eeq
Note that both $\psi_a$ and 
$\epsilon_{ab}$ have replica dimension $n \equiv (n_+ + n_-) \to 0$.
In this way we have for $G$
\beq
G(\mu,E) = \lim_{n_\pm \to \pm \mu/2\pi}
	\int D\overline{\psi} D\psi 
	\exp \left[ - \sum_{ab=1}^n \overline{\psi}_a \left(
	 M \delta_{ab}- i \epsilon_{ab} 
	\right) \psi_b \right] \ .
	\label{gg}
\eeq
The average of $G$ over the distribution of $J$ can be computed
by a generalization of the procedure of the previous section,
the main  difference being the fact that we have an extra replica 
dimension, so that the variable $q$ must be replaced by a matrix
$Q_{ab}$. For the sake of completeness the details of the computation
are in Appendix A.
We obtain
\beq
\overline{G(\mu, E)} = \int DQ \ e^{N S(Q,E)} \ ,
	\label{smodel}
\eeq
with
\beq
S(Q,E) = - \frac{1}{2} {\rm Tr} \, Q^2 + \log \det (- \hat E -iQ) \ ,
	\label{sq}
\eeq
and $\hat E_{ab}=
E \delta_{ab} + i \, \epsilon_{ab}$.
Note the similarity with equations (\ref{int1}) and (\ref{azione}).
The matrix $Q$ is an $n \times n$ 
self-dual real-quaternion matrix \cite{elk,meka2} (see Appendix A). 
It has $2 n^2 - n$ degrees of freedom, and is diagonalized
by transformations of the simplectic group $Sp(n)$.  
In (\ref{sq}) we see for the first time the role of
$\epsilon$ as a symmetry breaking field. The matrix $\hat E$ 
has an upper block of size $n_+$
which contains $+i\epsilon$ and a lower one of size $n_-$
with $-i\epsilon$, so that the action is only invariant under 
$Sp(n)/Sp(n_+) \times Sp(n_-)$, and the full invariance under
$Sp(n)$ is only recovered in the limit $\epsilon \to 0$. 
However, how {\it exactly} the symmetry breaking affects the calculation 
will become clearer below. 

We can evaluate the integral (\ref{smodel}) by means of the steepest 
descent, or saddle point, method, which becomes exact for large $N$. 
The saddle point equation for the matrix $Q$ reads
\[
Q = i (\hat{E} + i Q)^{-1} \ .
\]
This equation can be solved assuming for $Q$ a diagonal form, 
$Q_{ab} = z_a \delta_{ab}$.
We have two different sets of equations, one set for the elements
belonging to the upper block, $z_a^{(u)}$, 
and a second set for the elements
of the lower block, $z_a^{(l)}$. The only difference between the two 
sets is, of course, the sign of $\epsilon$,
\beqa
z_a^{(u)} &=& i \, \left( E + i \epsilon + i z_a^{(u)} \right)^{-1}
	 \mbox{upper block} \nonumber \\
z_a^{(l)} &=& i \, \left( E - i \epsilon + i z_a^{(l)} \right)^{-1}
	 \mbox{lower block}  
\eeqa
Each one of these two sets of equations has two solutions, $z_\pm^{(u)}$
for the upper block, $z_\pm^{(l)}$ for the lower one, namely
\beqa
z^{(u)}_\pm &=& 
	\frac{i}{2} (E + i\epsilon) 
	\pm 
	\frac{1}{2} \sqrt{4 - (E + i\epsilon)^2} \ ,
	\nonumber \\
z^{(l)}_\pm &=& 
	\frac{i}{2} (E - i\epsilon) 
	\pm 
	\frac{1}{2} \sqrt{4 - (E - i\epsilon)^2} \ .
\eeqa
For all values of the energy such that $|4-E^2| \gg \epsilon$
these solutions can be expanded in powers of $\epsilon$ and read
\beqa
z_\pm^{(u)} &=& q_\pm - \epsilon\left( 
	\frac{1}{2} \pm
	 i\,\frac{E}{2\sqrt{4-E^2}} \right) + O(\epsilon^2) \ ,
	\nonumber\\
z_\pm^{(l)} &=& q_\pm + \epsilon\left( 
	\frac{1}{2} \pm 
	i\,\frac{E}{2\sqrt{4-E^2}} \right) + O(\epsilon^2) \ , 
\label{speq}
\eeqa
where $q_\pm$ are given in equation (\ref{qpm}).

There are some important things to note here, related to the fact 
that the presence of $\epsilon$ crucially modifies the mutual 
relevance of the different saddle points. 
We have seen in the previous section that in the regime $-2 < E < 2$ 
the correct integration path $\gamma$ passes through both the saddle
points $q_+$ and $q_-$ (see Fig.2). 
This is true also in the present case, when a value $\epsilon\neq 0$ 
is considered: for each $z_a$ the path $\gamma$ passes through $z_+$ 
and $z_-$ and in principle both the saddle points must be taken into 
account.
However, when we look at the real part of the action $S$, 
we now discover that the contribution of one saddle point is 
exponentially dominant over the other by a factor
$\exp(-N\epsilon)$. This is in contrast with the case of the previous 
section, where the real part of $S$ was the same in the two 
saddle points.

The crucial point is that, due to opposite sign of $\epsilon$ in the 
upper and lower blocks, the real part of the action is tilted
in opposite ways in the two blocks and, as a consequence,
the dominant saddle point becomes $z_+$ for the upper block and 
$z_-$ for the lower one.
We now start to understand the way in which $\epsilon$ works
as a symmetry breaking field: without $\epsilon$ the two saddle points 
have the same weight in the integral and we have to consider both 
of them. With $\epsilon$, the weights are modified in 
opposite ways for the upper and lower blocks. 
In order to apply the steepest descent method we must perform
the limit $N\to\infty$ {\it before} the limit $\epsilon\to 0$, 
and this selects {\it just one} different saddle point for each 
of the two different blocks, dumping completely the non-dominant 
contribution. As a result, when at the end $\epsilon\to 0$ we have 
selected $q_+$ for the upper block and $q_-$ for the lower one. 
This is very reminiscent of what happens in statistical physics, 
where, in order to break a symmetry by means of an external 
field, the thermodynamic limit must be performed before sending  
the field to zero.

On the other hand, for energies $|E| > 2$, the effect of $\epsilon$ 
is  harmless,  there is no qualitative change from the situation
described in the previous section and  the same kind of
saddle point for the upper and lower block contributes to the integral.

We can now proceed in our computation.
We will focus first on the region $-2 < E < 2$, where the typical
spectrum is not positive defined and where we thus expect a more
interesting index distribution.
According to the above discussion on the dominant saddle points, 
we must consider the following form for $Q_{\rm SP}$:
\beq
Q_{\rm SP} = \mbox{diag}
	(
	\underbrace{z_+^{(u)}, \ldots, z_+^{(u)}}_{n_+}, 
	\underbrace{z_-^{(l)}, \ldots, z_-^{(l)}}_{n_-}
	)
	\label{sp} \ .
\eeq
This form is invariant under the unbroken group $Sp(n_+) \times
Sp(n_-)$ of replica symmetry transformations, and in this sense we
shall refer to it as a replica symmetric (RS) saddle point 
\cite{noiselle,meka1}.
We note that Eq.(\ref{dets}) is invariant under the simultaneous action
of complex conjugation and inversion of $\mu$, which after replicating
becomes $n_\pm \to n_\mp$, and that our saddle point 
satisfies this invariance.
If we plug expression (\ref{sp}) into Eq.(\ref{smodel}), we obtain 
after taking the limit $\epsilon\to 0$
\beq
\overline{G(\mu,E)} =
	\left(-E-iq_+ \right)^{N \mu/2\pi} \, 
	\left(-E-iq_- \right)^{-N \mu/2\pi} \, 
	e^{-N \, \mu \, (q_+^2-q_-^2)/4\pi} \,
	=
	\exp\left[i\mu N \alpha(E)\right] \ , 
\eeq
where $\alpha(E)$ is the average fraction of negative eigenvalues
given by Eq.(\ref{medio}). From (\ref{pi}) we finally get the
probability $p(k,E)$ in the limit $N \to \infty$,
\beq
p(k,E)= \delta \left[ k - \alpha(E) \right] \ .
	\label{delta}
\eeq
This result is very reasonable, but also rather trivial: the 
probability distribution of the intensive index is a $\delta$-function
peaked on its average value in the limit $N\to\infty$.
In order to observe a non-trivial behaviour we need to consider the
scaling with $N$, that is, the distribution of the index
for large but finite $N$. This is particularly
important if we are interested in the distribution of the 
extensive index, as for example in the case of disordered 
systems, where we want to know the change in the 
probability of different stationary points when variations 
of the index of order one, not of order $N$, are considered.

To go beyond result (\ref{delta}), we must consider
fluctuations around the saddle point (\ref{sp}). 
The general procedure is discussed in Appendix B.
As expected there are three kinds of fluctuations: within 
the upper block, within the lower block, and those
which mix the two blocks. 
Their corresponding eigenvalues and degeneracies are,
\beq
\begin{array}{lllcccl}
\omega_{\rm u} & = 
	1 + z_+^{(u)} \, z_+^{(u)} &
	=
	(1 + q_+^2) -
	\frac{\epsilon \, q_+^2}{\sqrt{1 - E^2/4}}
	+ O(\epsilon^2)& & & &
	d_{\rm u} = 2 n_+^2 - n_+ \ , 
	\\
\omega_{\rm l} & =
	1 + z_-^{(l)} \, z_-^{(l)} &
	=
	(1 + q_-^2) -
	\frac{\epsilon \, q_-^2}{\sqrt{1 - E^2/4}}
	+ O(\epsilon^2)& & & & 
	d_{\rm l} = 2 n_-^2 - n_- \ , 	
	\\
\omega_{\rm m} & =
	1 + z_+^{(u)} \, z_-^{(l)} &
	=
	\frac{\epsilon}{\sqrt{1 - E^2/4}}
	+ O(\epsilon^2) 
	& & & &
	d_{\rm m} = 4 n_+ n_- \ . \\
\end{array}
	\label{omega}
\eeq
The first two sets of eigenmodes are massive modes, in the sense that
their eigenvalues are $O(1)$. The third set are soft modes: 
for vanishing $\epsilon$ they would correspond to
zero modes associated to the restoration of the $Sp(n_+ + n_-)$
symmetry; for small non-zero
$\epsilon$ they become soft vibrations.
Integrating over the fluctuations, we obtain
\beq
\overline{G(\mu, E)} =
	\omega_{\rm u}^{-(n_+^2 - n_+/2)} \,
	\omega_{\rm l}^{-(n_-^2 - n_-/2)} \,
	\omega_{\rm m}^{-2 n_+ n_-} \,
	\exp\left[i\mu N \alpha(E)\right] \ .
\eeq
In the replica limit $n_\pm\to\pm\mu/2\pi$ this quantity becomes 
\beq
\overline{G(\mu, E)} =
	\exp\left[
	i\mu N \alpha(E) 
	- \frac{\mu^2}{2 \pi^2} \log 
	\left(
	\frac{\sqrt{\omega_{\rm u} \omega_{\rm l}}}{\omega_{\rm m}} 
	\right) 
	+ \frac{\mu}{4 \pi} \log 
	\left( \frac{\omega_{\rm u}}{\omega_{\rm l}}
	\right)
	\right] \ .
	\label{gege}
\eeq
From Eq.(\ref{pi}) we obtain the distribution 
for the extensive and intensive index for finite but large $N$:
\beqa
P(K,E) &=& \sqrt{\frac{1}{2 \pi \Delta(E)}} 
	\exp \left(
	- \frac{\left[ K - N \alpha(E) + \beta(E)\right]^2}
	{2 \Delta(E)}
	\right) \ , 
	\label{pK}
	\\
p(k,E) &=& \sqrt{\frac{N^2}{2 \pi \Delta(E)}} 
	\exp \left(
	- \frac{N^2\left[ k - \alpha(E) + \beta(E)/N \right]^2}
	{2 \Delta(E)}
	\right) \ . 
	\label{pk}
\eeqa
These are Gaussian distributions  peaked on the average
value  $\alpha(E)$. Indeed the shift, 
\beq
\beta(E)=\frac{1}{2\pi}{\rm arctg}\left(
\frac{E}{\sqrt{4-E^2}} \right) \ ,
\label{shift}
\eeq
is of order one and is not relevant at large enough values of $N$.
The variance $\Delta(E)$ is given by
\beq
\Delta(E) = \frac{1}{\pi^2} 
	\log 
	\left(
	\frac{\sqrt{\omega_{\rm u} \omega_{\rm l}}}{\omega_{\rm m}} 
	\right) \ ,
\eeq
that is
\beq
\Delta(E) = \frac{1}{\pi^2} 
	\log 
	\left[
	2 \pi^2 \epsilon^{-1} \rho_0(E)^2
	\right] \ ,
\label{var}
\eeq
where we have defined  $\rho_0(E) \equiv \rho(\lambda=0; E)$ 
(see Eq.(\ref{semi})).
This result for the variance can also be obtained by the method of
orthogonal polynomials where $\epsilon$ plays the role
of a high frequency cutoff \cite{ap2,ap4,review}. 

The fact that expression (\ref{var}) still depends on $\epsilon$ can
seem rather 
unphysical, especially when we consider the fact that the 
limit  $\epsilon\to 0$
has to be performed. However, we have to remember that we are looking
at finite $N$ corrections, and this very fact 
makes the parameters $\epsilon$ and $N$ no longer independent.
In this way the presence of $\epsilon$ translates in a
more physical $N$ dependence and this allows us to compute the 
scaling of the index distribution with the matrix size $N$.
Before discussing the result we have obtained for the index distribution, 
we have therefore to address the problem of the relation between $\epsilon$ 
and $N$.

There are mainly two different reasons why $\epsilon$ and $N$ are related.
First, as we have previously noted, there is a precise
interplay between the two limits, $N\to\infty$ and $\epsilon\to 0$,
when the saddle point approximation is used in order to solve integral 
(\ref{smodel}): the symmetry breaking due to $\epsilon$ works only if
$\epsilon\to 0$ {\it after} $N\to\infty$, as in any thermodynamic 
calculation. If $N$ is kept finite, we need a value of $\epsilon$ big
enough to guarantee the dominance of one saddle point over the other.
We have seen that the role of $\epsilon$ is to modify the real part of
the action in such a way that along the integration path $\gamma$ one
saddle point is weighted more than the other. However, if $\epsilon$
is too  small,  also the non-dominant saddle point may give a non-negligible
contribution to the  integral. To avoid this fact  we need 
the secondary contribution to be suppressed also at finite $N$ 
and to vanish when the limit $N\to \infty$ is considered.
The suppression factor is given, at order $\epsilon$,  by
\beq
e^{- N [S(z_+^{(u)}) - S(z_-^{(u)}) ]} = e^{- 2 \pi N \epsilon \rho_0(E) } \ ,
\label{supp}
\eeq 
for the upper block (for the lower block an analogous expression is valid).
In order for the suppression factor to vanish it must hold
\beq
\epsilon N \to \infty \ , \ \ \ \ \ \ \ \ N \to \infty \ .
	\label{grande}
\eeq
This imposes a lower bound for $\epsilon$ when $N$ is finite. 
A natural general choice is therefore to assume
\beq
\epsilon=\frac{1}{N^{1-\delta(N)}} \ ,
\label{scale}
\eeq
where the exponent $\delta(N)$ has to satisfy the relation 
$\delta(N) \log N \to \infty$. The simplest possibility is, of course, a
constant value of $\delta$. However, as we shall argue immediately
below, this would not be consistent with the second condition we have to
impose on $\epsilon$. 

The second bound for $\epsilon$ comes from the following observation.
When we perform our calculation with a finite value of $N$ and of $\epsilon$, 
there are of course two different kinds of corrections to the 
asymptotic exact result:
the first kind is related to the saddle point approximation and brings
corrections which scale as inverse powers of $N$. The
second is related to the non-zero value of $\epsilon$ and brings corrections
which scales as powers of $\epsilon$. Consistency requires that in the final 
result the error introduced by considering a finite value of $\epsilon$
must be of the same order as the terms we discard in the expansion in
$1/N$.
It can be easily shown that the corrections to
the index distribution (\ref{pK}) for finite $\epsilon$ are of order 
$\epsilon^2$, that is 
\beq
G(\mu)=\exp(N\alpha(E)+ \beta(E) + O(\epsilon^2)) \ .
\label{lbound}
\eeq
On the other hand, by considering the Gaussian fluctuations around the
saddle point, we are discarding terms of order $1/N^2$ in the exponent 
of (\ref{lbound}). Thus, we must impose the condition
\beq
\epsilon^2 \sim \frac{1}{N^2} \ .
\label{chico}
\eeq
Equation (\ref{chico}) is consistent with equations (\ref{scale}) and
(\ref{grande}) only if,
\beq
\delta(N)\to 0    \ , \ \ \ \ \ 
\delta(N)\log N\to \infty \ , \ \ \ \ \ \
N\to\infty \ .
	\label{patty}
\eeq
In this way we finally get for the variance the result,
\beq
\Delta(E,N) = \frac{1}{\pi^2} 
	\log 
	\left[
	4 \pi^2 N^{(1-\delta(N))} \rho_0(E)^2
	\right] =
	\frac{1-\delta(N)}{\pi^2}\log N
	+ \frac{2}{\pi^2}\log(2\pi\rho_0) \ ,
\label{varn}
\eeq
where we have taken $\epsilon=1/2N^{1-\delta}$, the factor $1/2$ 
being consistent with equation (\ref{supp}) at $E=0$.
This result agrees very well with numerical simulations: in Fig.3 we plot
the variance $\Delta$ as a function of $\log N$, obtained by exact
numerical diagonalization. A linear fit gives
\beq
\Delta = \frac{a}{\pi^2} \log N + \frac{b}{\pi^2} \log(2\pi\rho_0) \ ,
\ \ \ \ \ a=1.005\pm 0.006 \ \ , \ \ b= 1.993\pm 0.003 \ .
\eeq
This same scaling for the variance has been found also in
\cite{ap2}, where a completely different method based on the
invariance properties of the Gaussian Orthogonal Ensemble and the
dominance of intrinsic binary correlations was used.
In Appendix B we show in details that the contributions of the other possible 
saddle point solutions of the whole integral (\ref{smodel}) to the index
distribution are smaller by inverse powers of $\log N$ in this energy
region, therefore the scaling with $N$ is correctly reproduced by equations
(\ref{pK}),(\ref{pk}) and (\ref{varn}).

\begin{figure}
\begin{center}
\leavevmode
\epsfxsize=3in
\epsffile{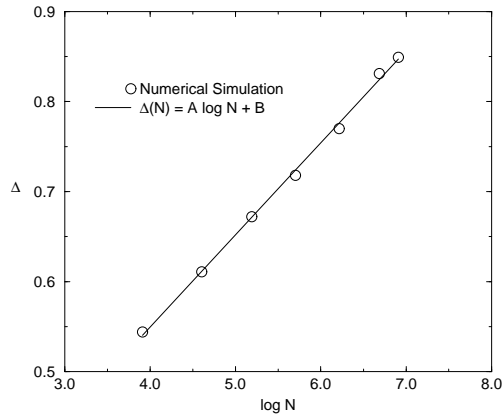}
\caption{The variance $\Delta$ as a function of 
$\log N$ for $E=0$, obtained by means of exact numerical diagonalization on
the Gaussian Orthogonal Ensemble. The full line is the linear fit.}
\label{fig3}
\end{center}
\end{figure}

We can finally analyze the significance of our result, equations 
(\ref{pK}),(\ref{pk}) and (\ref{varn}), in the energy regime
$-2<E<2$.
What we see is that the variance of the probability distribution 
of the {\it intensive} index goes to zero in the limit $N\to\infty$
and this was quite expected, given our former result (\ref{delta}).
On the other hand, the variance of the distribution of the 
{\it extensive} index diverges logarithmically for $N \to \infty$.
The meaning of this result is the following: 
on the one hand the probability of finding a matrix with an 
index density different from the average one, that is with an extensive 
index ${\cal I} \sim N\alpha + O(N)$, is zero in the limit
$N\to\infty$. But, on the other hand, the probability of having a matrix
whose index differs from the average one for a number of
negative eigenvalues of order one, i.e. ${\cal I} \sim N\alpha + O(1)$,
is exactly the same as the probability of having a matrix with the average 
index, in the limit $N\to\infty$. 
As we shall see in the next section, this fact has
some very interesting physical consequences in the context of 
disordered systems.

Let us now look at the other energy regions.
First of all we note that the derivation of equations
(\ref{speq}), (\ref{omega}) and (\ref{supp})
holds as long as the energy is such that $\rho_0(E)$ is of $O(1)$. 
But this condition breaks down  when the energy gets close to $\pm 2$
and $\rho_0(E) \ll 1$. In this region the procedure previously adopted  
to compute the index distribution has to be modified. 
Indeed, when $\rho_0(E)$ becomes too small the suppression mechanism 
(\ref{supp}) starts being inefficient, and the saddle point (\ref{sp}) is no
longer  the only one contributing to the integral. 
At some point the excitations which were treated as soft modes in
(\ref{omega}) must be considered  as zero modes connecting  equivalent
saddle points: there exists a manifold $Sp(n_+ + n_-) / Sp(n_+) \times
Sp(n_-)$ of saddle points and the original replica symmetry under $Sp(n_+ +
n_-)$ is restored. 
At this stage $\epsilon$ plays no longer any role
and it can be taken to zero. 
The massive modes are the same as in 
Eq. (\ref{omega}), and after integrating over them and exactly over
the degrees of freedom associated with the zero modes we obtain
(up to trivial factors in the replica limit)
\beq
\overline{G(\mu, E)} =
	N^{2 n_+ n_-}
	{\cal V}^{n_+}_{n_+ + n_-} \,
	\omega_{\rm u}^{-(n_+^2 - n_+/2)} \,
	\omega_{\rm l}^{-(n_-^2 - n_-/2)} \,
	\exp\left[i\mu N \alpha(E)\right] \ ,
\eeq
where ${\cal V}^{n_+}_{n_+ + n_-}$ corresponds to the volume of the 
manifold of saddle points solutions (see Appendix B). 
At this point one has to analytically continue the previous
expression for $n_{+} \to \mu/2\pi , \, n \to 0$. 
The volume ${\cal V}^{n_+}_{n_+ + n_-}$ is finite for $n_+=0$ and
it is zero for positive integers \cite{meka1}. Its analytic
continuation is an oscillatory function of $n_+$, with exponentially
increasing amplitude \cite{zirn}, so that the presence of this factor in 
the former equation makes the index distribution non-Gaussian.
However, as long as this analytic continuation is finite for 
non-integer $n_+\sim 1$, the distribution can be approximated by 
a Gaussian with variance 
\beq
\Delta(E \sim \pm 2) = \frac{1}{\pi^2} 
	\log 
	\left[
	8 \pi^4 \, N \, \rho_0(E)^3
	\right] \ .
\label{delta2}
\eeq
We can see from (\ref{delta2}) that the variance still scales as
$\log N$. However,  when $|E-2| \sim 1/N^{\frac{2}{3}}$, we have 
$\rho_0(E)\sim 1/N^{\frac{1}{3}}$ and a 
further crossover takes place: the variance $\Delta(E)$ becomes of
order one  meaning that the index distribution is dramatically
more peaked around its typical value as we approach  $E=\pm 2$.
Note also that when  $E \sim -2 + 1/N^{\frac{2}{3}}$ the typical index
$\alpha(E)$ becomes of order 1, meaning that in this region matrices
with $O(1)$ negative eigenvalues are dominant.
Summarizing, in the energy regime where the average number of negative 
eigenvalues is of order one, the fluctuations around the mean value
become of order one too.

When the energy is exactly at the threshold values  $E = \pm 2$
we have a special case since the saddle point equations for the eigenvalues have
a single degenerate solution, and the harmonic terms in the expansion
around the saddle point vanish. It is not difficult to show that 
the distributions here become
\beq
P(K,E \to -2^+) = N^{-1} \, \delta(K)  \ , 
	\;\;\;\;\;\;\;
P(K,E \to 2^-) = N^{-1} \, \delta(K - N) \ .
	\label{limit}
\eeq

The calculation in the regions $|E| > 2$ is completely straightforward
since $\epsilon$ plays no role from the beginning.
As mentioned before, the same kind of saddle point 
has to go in both blocks, so that we have
\beq
Q_{\rm SP} = \mbox{diag}
	(
	\underbrace{z_\pm^{(u)}, \ldots, z_\pm^{(u)}}_{n_+}, 
	\underbrace{z_\pm^{(l)}, \ldots, z_\pm^{(l)}}_{n_-}
	)
	\label{spout} \ ,
\eeq
where the plus (minus) sign corresponds to negative (positive)
energies. There is only one kind of massive fluctuation with
degeneracy $2 n^2 - n$, which goes to zero in the replica
limit, and thus the integration over fluctuations gives a trivial
prefactor. The final result for the distribution of $K$ is
\beq
P(K,E) = \left\{
	\begin{array}{lc}
	N^{-1} \, \delta(K)  & E < -2 \\
	N^{-1} \, \delta(K - N)  & E > 2 \ , \\
	\end{array}
	\right. 
\label{below}
\eeq
which coincides with the limiting behaviour (\ref{limit})
of the distribution in the region $-2<E<2$. Thus, while in the
energy region $-2<E<2$ values of the index with an $O(1)$ difference from
the typical one have a finite probability, here the index distribution
is so much peaked on the typical value that even small changes in the
index have zero probability.

\section{An application to disordered systems}

In this section we consider a mean-field spin-glass model, that
has been extensively studied in the last years and whose thermodynamical as
well as dynamical features are very well known, namely the $p$-spin 
spherical model \cite{crisa1,crisatap,crisa2,kpz,ck1}. 
Our aim is to use the results of the calculation we have carried out 
in the previous section, in order to have a better understanding of the 
statistical and geometrical properties of the energy landscape for this 
model.

This problem is by itself relevant, because both the static
properties and the peculiar off-equilibrium dynamical behaviour 
of mean field spin-glasses, and in particular of this model,  
are known to be deeply related to the distribution of the minima 
and of the saddles of the Hamiltonian \cite{spin-glass,laloux,noiselle,nomo}.
Moreover, it is now commonly accepted that the $p$-spin spherical model
shares many common features with structural glasses, which are presently 
one of the major challenges for statistical mechanics.  Indeed, 
notwithstanding the completely different form of the Hamiltonians, 
some structural glasses (in particular fragile 
glasses) and the $p$-spin spherical model have a very similar 
structure of the energy landscape \cite{francesi}. 
Therefore, a thorough investigation of the energy landscape for
the $p$-spin spherical model is  important also for a better
understanding of structural glasses. 

As already stated in the Introduction, knowing the index 
distribution of the Hessian at various energies is equivalent to knowing 
the fluctuations in the stability of the energy surface. In other words,
the index distribution tells us what are the dominant stationary points
of the Hamiltonian (or saddles) at a given energy, and, more importantly, 
what is the probability distribution around the typical saddles, 
thus providing an insight on the mutual entropic accessibility of different 
stationary points. This is what we are going to describe in this last section.

The reason why the $p$-spin spherical model is particularly
appropriate for an application of the above calculation and concepts 
is the following: when we look at the stationary points of the
Hamiltonian of this system, we find that the Hessian matrix $M$ in such 
stationary points, behaves as a Gaussian
random matrix of the same kind as the ones considered in the
calculations above. More specifically, if we classify the stationary 
points of the Hamiltonian in terms of their energy density $E$, 
we find that the Hessian $M(E)$ in these stationary points is
a random matrix of the form (see for instance \cite{kpz})
\beq
M_{ij}(E) = J_{ij} - E \, \delta_{ij},
\eeq
where $J_{ij}$ is an $N$-dimensional real and symmetric random matrix 
with the same Gaussian distribution as (\ref{dist}), and where $N$ is 
the size of  the system. The spectrum of the Hessian in the stationary
points is therefore, 
\beq
\rho(\lambda;E)= 
	\frac{1}{2\pi} \sqrt{E_{th}^2 - (\lambda+E)^2} , 
\eeq
where $E_{th}$ is the so-called {\it threshold energy}, which depends
on the parameters of the model (in the previous sections it was 
$|E_{th}|=2$).
Given the particular shape of the Hessian, we can completely
disregard the details of the $p$-spin spherical model and
assume the results obtained in our calculation 
as the starting point, interpreting these results 
in terms of probability distributions of the stationary points of
the Hamiltonian.

Let us begin our geometric analysis of the energy landscape from very low
energies. When $E<-|E_{th}|$ the semi-circle is entirely contained in the
positive semi-axis and the average determinant of the Hessian is positive: 
this is the region dominated by minima, as the index distribution (\ref{below})
shows. Moreover, as we have already noted in the previous section, 
the probability of finding a stationary point with an index different 
from the typical one (i.e. $0$) is zero. Minima are strongly dominant
in this energy regime. 
A more careful analysis \cite{noiselle} shows that even in 
this regime there are saddles with non-zero index, 
but the probability of these objects is exponentially small in $N$, that is 
\beq
P(K,E) \sim e^{-KN\Omega(E)} \ \ \ \ \ \ K=1,2,3, \dots \ .
\eeq
This result is obtained by considering non-symmetric contribution to the 
saddle point equations (see \cite{noiselle} and Appendix B) and,
consistently with equation (\ref{below}), it gives a contribution too
small to be caught by simply analyzing  fluctuations around the
dominant saddle point.  
The above result shows that at low energies minima are exponentially 
dominant over saddles of order one, and even more dominant over saddles
with extensive index. In this sense we shall call this region the 
{\it decoupling regime},  since at any energy below the threshold only
one kind of stationary points, namely minima, dominates. 
When we raise the energy, we finally arrive at $E=-|E_{th}|$: here the 
semi-circle touches the zero and the decoupling between different
stationary points is no longer true. Indeed, it can be proved \cite{noiselle}
that $\Omega(E_{th})=0$,  meaning that at the threshold energy
minima and saddles of order one have the same probability.

Thanks to the calculation of the previous section we are now in the position 
to answer the following question: What happens {\it above} the threshold
energy ? 
From a simple inspection of the semi-circle law it is  clear that
above the threshold saddles become important, since many negative eigenvalues
appear and the average index $N\alpha(E)$ is non-zero. Yet, in order to have 
information on the degree of decoupling of the stationary points, the simple 
typical index  $N\alpha(E)$ is not enough. The reason is the following:
the knowledge of the typical index does not tell us whether at that same 
energy other stationary points, different from the typical ones, do
or do not have non-zero probability. In this sense the mutual entropic
accessibility of different stationary points is encoded in the 
index distribution $P(K,E)$, which reveals to what extent the typical
saddles are dominant over the non-typical ones.

From  equation (\ref{pK})  we see that in this regime not 
only the dominant stationary points
are saddles of order $N$, but, also, that the probability of finding a minimum 
is of order $e^{-N^2}$. The decoupling between minima and dominant saddles
is therefore much more dramatic than the one we found below 
the threshold. 
On the other hand, because of the divergence of the variance $\Delta$ with $N$ 
(equation (\ref{varn})), we see that there is a mixing among saddles
with the same {\it intensive} index: the probability of having a saddle 
whose index differs from the average by a number of order one, 
is the same as the probability of the 
typical saddles \cite{nota}. In other words, the main result is 
that there is no decoupling among saddles with the same intensive index,
so that a mixing of different stationary points occurs, while still a
decoupling exists between dominant saddles and minima.

Summarizing, we can therefore distinguish two  energy regimes
where the probability distribution  of the stationary points, 
and therefore the geometric structure of the energy landscape, 
is very different: a decoupled regime for $E<E_{th}$ and a mixed
regime for $E>E_{th}$.
Interestingly enough, the threshold energy $E_{th}$ is exactly the 
asymptotic energy where a purely dynamical transition
occurs: below a critical temperature $T_d$, the ergodicity 
is broken and the system is no longer able to visit the entire phase 
space in its time evolution, remaining confined to an energy level 
higher than the equilibrium one. This `dynamical energy' is equal to 
$E_{th}$ \cite{crisa2,kpz,ck1}.

This suggests us to relate the information we have on the distribution 
of the stationary points, following from the index distribution, to the 
dynamical physical behaviour of the system. 
Above $T_d$ the equilibrium energy $E$ of the system is higher than 
threshold value $E_{th}$ and therefore belongs to the mixed regime:
the equilibrium landscape explored by the  system is dominated by  
saddles of order $N$ which, as we have shown, are all equally relevant up
to variations of the index of order one. This means that all these unstable
stationary points are equally accessible 
to the system in its time evolution.
As $T_d$ is approached the equilibrium energy $E$ gets closer and 
closer to $E_{th}$, and the properties of the equilibrium landscape 
change accordingly to the behaviour of $P(K,E)$ we have discussed in
the previous section: when $E\sim -|E_{th}| + 1/N^{2/3}$ 
saddles with index of order one become the most relevant and 
the variance of the index distribution is now finite. This means that
minima start having a finite probability in this energy regime. 
The range of temperatures where this behaviour takes places is of order 
$1/N^{2/3}$ and shrinks to zero in the thermodynamic limit. Below $T_d$, 
the equilibrium energy belongs to the decoupled regime, that is $E<E_{th}$: 
minima are now dominant and saddles of any order have exponentially 
vanishing probability.
We can therefore interpret $T_d$ as the temperature where a 
geometric transition occurs from a regime of strong mixing of the 
stationary points to a regime of equally strong decoupling.

\section{Conclusions}

In this paper we computed the average index distribution for an
ensemble of Gaussian random matrices. We find a result which is
in optimum agreement with exact numerical diagonalization.
This computation is, in our opinion, an interesting example where the
fermionic replica method, together with a careful asymptotic expansion
of the integrals, gives correct results. We hope that the present 
work can therefore contribute to clarify the role of 
the replica method in the context of RMT.

Besides, and this was our main purpose, the index distribution
provides a really useful tool for investigating the geometric
structure of the energy landscape in disordered systems. 
In the previous section we applied this tool to the
simple case of the $p$-spin spherical model and discussed the physical
consequences of our results. In general, the task of computing the
distribution of the index of the Hessian 
is not as simple as in the $p$-spin model. The main
reason is that the Hessian usually does not  
behave as a Gaussian random matrix,
because, as noted in the Introduction,  
its distribution is determined both by the distribution of the
quenched disorder and by the distribution of the configurations. 
However, the same procedure we adopted in this
paper can also be applied to these more complicated cases, with the
appropriate modifications: to compute the index
distribution at a given energy $E$, 
one has to average over the distribution of the disorder and integrate
over the relevant configurations belonging to
the manifold of energy $E$ \cite{noiselle}.
This is the reason why the method presented in this paper
is particularly suitable for this task,
since it addresses the problem without 
assuming any particular form for the 
distribution of the Hessian.

Finally, there have been recently some attempts to find a connection
between the occurrence of a thermodynamical phase transition and the 
change in the topology of the configuration space visited by the system at
equilibrium \cite{lapo1}. 
For various non-disordered models which present a second order phase 
transition it has been shown via numerical simulations that the
fluctuations of the curvature of the configuration space exhibit a
singular behaviour at the transition point. 
This is  similar to the behaviour described in the previous section
for the $p$-spin spherical model, where the average fluctuations of the
index (\ref{var}) at the equilibrium energy encounter a dramatic
change as the dynamical transition is approached \cite{noiselle}. 
This suggests first of all that also in disordered systems a connection
between thermodynamical behaviour
and topology of the configuration space exists. Besides, the case of
the $p$-spin is also peculiar in this sense: 
it presents a static phase transition which is thermodynamically of
second order, but it is discontinuous in the order parameter
\cite{crisa1} and exhibits a purely dynamical transition at a higher
temperature \cite{crisa2}. As we have shown, in this case a dramatic
change of geometrical  properties occurs at the dynamical
transition, indicating that a  more complex situation probably holds 
for disordered systems which present this sort of behaviour.

\acknowledgements
It is a pleasure to thank Jorge Kurchan for many suggestions 
and discussions. We also thank Kurt Broderix for a careful reading of
the manuscript and for some key remarks on a preliminary version
of this work. Finally, we wish to thank
the kind hospitality of the Department of Physics of the ENS of Lyon, 
where part of this work was done. A.C. and I.G were supported by
EPSRC Grant GR/K97783, and  J.P.G. by EC Grant
ARG/B7-3011/94/27.

\appendix

\section{}
In this Appendix we give for completeness the standard procedure
used to average and integrate out the fermions which gives the 
sigma model (\ref{smodel})-(\ref{sq}) \cite{elk}.

The average over the Gaussian Orthogonal Ensemble (GOE) of Eq.(\ref{gg}) yields,
\beq
G(\mu,E) = \lim_{n_\pm \to \pm \mu/2\pi}
	\int D\overline{\psi} D\psi 
	\exp 
	\left(
	\hat{E}_{ab} \, \overline{\psi}_a \cdot \psi_b
	+ \frac{1}{2 N} 
	\overline{\psi}_a \cdot \overline{\psi}_b \, \psi_b \cdot \psi_a 
	- \frac{1}{2 N} 
	\overline{\psi}_a \cdot \psi_b \, \overline{\psi}_b \cdot \psi_a 
	\right) ,
\eeq
where summation over repeated replica indices is implicit and 
the dot stands for contraction of spatial indices. We can define
the following $n \times n$ matrix whose components are quaternions,
\beq
A_{ab} = 
	A^0_{ab} \, {\bf 1} +
	\sum_{s=1}^3 A^s_{ab} \, {\bf e}_s ,
\eeq
where
\beqa
A_{ab}^0 &=& \frac{1}{4} \left( 
	\opsi_a \cdot \psi_b + \opsi_b \cdot \psi_a 
	\right) ,
	\\
A_{ab}^1 &=&	
	\frac{i}{4} \left( 
	\opsi_a \cdot \opsi_b - \psi_a \cdot \psi_b 
	\right) ,
	\\
A_{ab}^2 &=&	
	- \frac{1}{4} \left( 
	\opsi_a \cdot \opsi_b + \psi_a \cdot \psi_b 
	\right) ,
	\\
A_{ab}^3 &=&
	\frac{i}{4} \left( 
	\opsi_a \cdot \psi_b - \opsi_b \cdot \psi_a 
	\right) ,
\eeqa
and $\{ {\bf 1}, {\bf e}_1, {\bf e}_2, {\bf e}_3 \}$ are the basis
for the field of quaternions \cite{mehta}, which can be represented
by two by two matrices,
\beq
{\bf 1} = \left( \begin{array}{cc} 1 & 0 \\ 0 & 1 \end{array} \right) ,
	\;\;\;\;\;
{\bf e}_1 = \left( \begin{array}{cc} 0 & -i \\ -i & 0 \end{array} \right) ,
	\;\;\;\;\;
{\bf e}_2 = \left( \begin{array}{cc} 0 & -1 \\ 1 & 0 \end{array} \right) ,
	\;\;\;\;\;
{\bf e}_3 = \left( \begin{array}{cc} -i & 0 \\ 0 & i \end{array} \right) .
\eeq
Every $n \times n$ quaternion matrix $A$ can be represented
by a  $2n \times 2n$ complex matrix $C(A)$, and the following
relations
hold: $\det^2 A = \det C(A)$ and $2 \, {\rm Tr} \, A = {\rm Tr} \, C(A)$, where
for the quaternion matrix the trace selects the real part (component
of {\bf 1}) at the end. 
Using these properties, 
the quartic terms in the fermions can then be written as
\beq
\frac{1}{2N} 
	\overline{\psi}_a \cdot \overline{\psi}_b \, \psi_b \cdot \psi_a 
	- \frac{1}{2N} 
	\overline{\psi}_a \cdot \psi_b \, \overline{\psi}_b \cdot \psi_a 
	= - \frac{1}{2N} {\rm Tr} A^2 ,
\eeq
where the trace also selects the scalar part of the quaternion. 
These quartic terms  can be decoupled by a Hubbard-Stratonovich 
transformation, 
\beq
\exp - \frac{1}{2N} {\rm Tr} A^2 =
	\int DQ \exp \left( - \frac{N}{2} {\rm Tr} \, Q^2 + 
	i {\rm Tr} \, A \, Q \right) ,
\eeq
The matrix $Q$ is again a quaternion $n \times n$ matrix,
\beq
Q_{ab} = 
	Q^0_{ab} \, {\bf 1} +
	\sum_{s=1}^3 Q^s_{ab} \, {\bf e}_s ,
\eeq
with $Q^0$ real and symmetric and $Q^s$ real and antisymmetric,
so that $Q$ has $2n^2 -n$ degrees of freedom. 
Such matrices are called self-dual real quaternion. The fermions
can now be integrated out, and we obtain
Eqs.(\ref{smodel})-(\ref{sq}).

\section{}
In this Appendix we calculate the contributions to the index
distribution of saddle point (SP) solutions different from Eq. (\ref{sp})
for energies in region I. 

A general SP solution reads,
\beq
Q_{\rm sp} = \mbox{diag}
	\left(
	\underbrace{z_+^{(u)},\ldots,z_+^{(u)}}_{p_+}, 
	\underbrace{z_-^{(u)},\ldots,z_-^{(u)}}_{n_+ - p_+},
	\underbrace{z_-^{(l)},\ldots,z_-^{(l)}}_{p_-}, 
	\underbrace{z_+^{(l)},\ldots,z_+^{(l)}}_{n_- - p_-}
	\right) .
	\label{eq:sp}
\eeq
For $p_\pm \neq 0, n_\pm$ these SP are not invariant under the
unbroken replica symmetry $Sp(n_+) \times
Sp(n_-)$ of Eq. (\ref{smodel}), 
and have therefore been called replica symmetry broken (RSB)
solutions \cite{meka1}, 
even though the symmetry is subsequently restored by zero modes.
The action (\ref{sq}) at the saddle-point
solution (\ref{eq:sp}) reads, after taking the replica limit,
\[
i N \alpha(E) \left[ \mu - 2 \pi (p_+ - p_-) \right] .
\]
Lets consider now the fluctuations \cite{meka2}.
There are sixteen different normal modes. 
They are labeled by the pair
of indices $(\alpha, \sigma)$, where the index $\sigma = \pm 1$ indicates
the upper and lower block, and the index $\alpha = \pm 1$ indicates 
the sub-block with solution $z_{\alpha}$.
The eigenvalues are thus denoted by 
$\omega_{(\alpha, \sigma)(\alpha',\sigma')}$.
There are three kinds of fluctuations:

\begin{description}
\item (i) Massive modes: these correspond to $\alpha = \alpha'$ for any 
$\sigma$ and $\sigma'$, with eigenvalues
$\omega_{(\alpha \sigma)(\alpha \sigma')} = (1 + q_{\alpha}^2)$. 

\item (ii) Zero modes: for $\alpha = - \alpha'$ and $\sigma = \sigma'$.
They are present for any RSB solution.

\item (iii) Soft modes: for $\alpha = - \alpha'$ and $\sigma = -
\sigma'$,
with eigenvalue $\omega_{(\alpha \sigma)(-\alpha -\sigma)} = \alpha
\sigma \epsilon / \pi \rho_0(E)$. 
\end{description}

Lets consider 
SP's with $p_+ = p_- = p$, which respect the symmetry of the problem
under simultaneous complex conjugation and inversion of $\mu$.
For these, after integrating out the fluctuations, we obtain
\beq
\overline{G(\mu, E)}_{p_\pm = p} =
	\lim_{n_\pm \to \pm \mu/2\pi}
	\, {\cal V}_{n_+}^{p} 
	\, {\cal V}_{n_-}^{p}
	\, \left[ \pi \rho^2_0(E) \right]^{4 p^2}
	\, \overline{G(\mu, E)}_0 ,
	\label{eq:gig}
\eeq
where $\overline{G(\mu, E)}_0$ is the result from the RS solution
(\ref{gege}) (with  $\epsilon \sim 1/N^{1-\delta}$, 
see Eq.(\ref{patty})) and  the volume of the
zero mode manifolds ${\cal V}_{n_\pm}^p$ is given by
\beq
{\cal V}^p_n = 
	\left[ 4 \pi^3 \, \rho_0^2(E) \right]^{2 p (n-p)}
	\, F_n^p ,
	\;\;\;\;\;\;\;
F^p_n = 
	\frac{\Gamma(1+n)}
	{\Gamma(1+p) \, \Gamma(1+n-p)} \, 
	\prod_{j=1}^p \,
	\frac{\Gamma(1+2j)}{\Gamma[1 + 2(n-j+1)]} .
	\label{vol}
\eeq

We need now to determine the zero-modes volume in
the replica limit. Using the property of the Gamma function $\Gamma(z)
\Gamma(1-z) = \pi / \sin \pi z$, and noting that $n_\pm \to \pm
\mu/2\pi \sim N/\Delta(E)$, so we want the large $|n_\pm|$ limit,
we find that
\beq
F_{n_+}^{p} \to
	\left(
	\frac{(-1)^{p^2} \pi^{-4p^2}}{\Gamma^2(1+p)} \, 
	\prod_{j=1}^p \Gamma^2(1+2j) 
	\right) \, 
	\left( \frac{\mu}{2 \pi} \right)^{2 p^2 -p} \, 
	\sin^p \mu .
	\label{eq:vol}
\eeq
We can now use Eq. (\ref{pi}) to obtain the contribution to the index
distribution of the RSB solutions. For the simplest one $p=1$ we
get,
\beqa
P_1(K,E) &=& 
	\left[ 4 \pi^2 \, \rho^2_0(E) \right]^{-4}
	\, \sqrt{\frac{2}{\pi \, \Delta^3(E)}}
	\, \left[ 
	\left[ K - N \alpha(E) + 1 \right]
	\,
 	\exp \left(
	- \frac{\left[ K - N \alpha(E) + 1 \right]^2}{2 \Delta(E)}
	\right) 
	\right.
	\non \\
	&&
	- 
	\left.
	\left[ K - N \alpha(E) - 1 \right]
	\,
 	\exp \left(
	- \frac{\left[ K - N \alpha(E) - 1 \right]^2}{2 \Delta(E)}
	\right) \right] .
	\label{eq:pk1}
\eeqa
For $|E| < 2$ this is an $O(1/\log N)$ contribution to the
distribution (\ref{pk}) obtained from the RS solution. 
In contrast to the case of correlation functions \cite{meka1},
contributions from 
higher RSB saddle-points do not vanish, but give
contributions decreasing by powers of $\log N$.

Lets turn now to the external regions $|E| > 2$. The RSB solutions
here are those with $p_+ = (n_- - p_-) = p > 0$. Evaluating 
(\ref{sq}) in these SP's we find that their contributions are
suppressed by
\[
\exp\left\{ - p N \left[ 
	|E| \sqrt{E^2 - 4} + 
	\log \left( \frac{E^2}{2} + 
	\frac{|E|}{2} \sqrt{E^2 - 4} + 1 \right) \right] \right\} ,
\]
as was already found in \cite{noiselle}.


\begin{thebibliography}{}
\bibitem{Jorge}
We wish to thank Jorge Kurchan for suggesting us the use of this formula 
and its representation in terms of replicated fields (see equation (\ref{gg})).
\end{thebibliography}
\end{document}